\documentclass{article}
 \def\be{\begin{equation}}      
  \def\ee{\end{equation}}    \def\beq{\begin{eqnarray}}
  \def\dis{\displaystyle}    \def\eeq{\end{eqnarray}}
       \def\m{\multicolumn}
         \def\ra{\rightarrow}
\begin{document}

\begin{center}
{\large\bf Decay rates of quarkonia with NRQCD formalism using spectroscopic parameters of potential models}

{\em Ajay Kumar Rai$^1$, J N Pandya$^2$, P C Vinodkumar$^3$}

{$^1$Physics Section, Applied Sciences and Humanities Department, Sardar Vallabh National Institute of Technology, Surat 395007, Gujarat, INDIA. \\ $^2$Applied Physics Department, Faculty of Technology \&
Engineering, The M S University of Baroda, Vadodara 390001, Gujarat, INDIA. \\ $^3$Department of Physics, Sardar Patel University, Vallabh Vidyanagar 388 120, Gujarat, INDIA.}

\end{center}

{\bf Abstract:}\\ Decay rates of quarkonia are studied within the framework of NRQCD formalism. The basic parameters of the formalism have been obtained from different potential schemes studied for the spectra of quarkonia. We estimate the heavy quarkonia mass spectra, radiative and leptonic widths and compare them with other contemporary theoretical approaches and experimental results.

\baselineskip=16pt
\section{Introduction}
\label{intro}
The spectroscopy and decay rates of quarkonia are quite important to study due to huge amount of high precession data acquired using number of experimental facilities at BES at the Beijing Electron Positron Collider (BEPC), E835 at Fermilab and CLEO at the Cornell Electron
Storage Ring (CESR), the B-meson factories, BaBar at PEP-II, Belle at KEKB, the CDF and D0 experiments at Fermilab, the Selex experiment at Fermilab, ZEUS and H1 at DESY, PHENIX and STAR at RHIC, and NA60 at CERN \cite{nb2007}. New states and production mechanisms, new decays and transitions have been identified and even larger data samples are expected to come from the BES-III upgraded experiment, while the B factories and the Fermilab Tevatron will continue to supply valuable data for few years \cite{nb2007}. New facilities like the LHC experiments at CERN, Panda at GSI etc. will offer greater challenges and opportunities in this field \cite{nb2007,M.B.Voloshin2008,Eichten2008}.

The mesonic states are not only identified with their masses but also with their various decay rates. So, one of the tests for the success of any theoretical model for mesons is the correct prediction of their decay rates. Many phenomenological models predict the masses correctly but overestimate the decay rates \cite{BuchmullerTye1981,Martin1980,Quiggrosner1977,Eichten1978,AKRai2002,AKRai2005,AKRai2006,Gerstein1995}.
 For better estimates of their decay rates with reference to the experimental values, various corrections due to radiative processes, higher order QCD contributions etc have been suggested \cite{Hafsakhan1996}. In this context, the NRQCD formalism is found to provide systematic treatment of the perturbative and non-perturbative components of QCD at the hadronic scale \cite{Bodwin1995,Bodwin2006,Bodwin2008}. For the present study, we employ phenomenological potential schemes for the bound states of heavy quarkonia and the resulting parameters and wave functions to study the decay properties. For example, we include here the extended relativistic harmonic confinement scheme \cite{JNPandya2001,PCV1999} employed for computation of quarkonia masses and decay rates along with other potential approaches such as coulomb plus power potential (CPP$_\nu$) with different choices of the power index $\nu$ etc \cite{AKRai2005} to study the meson properties.
\section{Potential model approaches for quarkonia spectroscopy}
\label{sec:1}
In this section, we review important features of the models such as ERHM and coulomb plus power potential (CPP$_\nu$) employed in the spectroscopic study of heavy flavour hadrons. We extract the spectroscopic parameters of the models such as the model parameters that reproduces the quarkonium spectra, the wave function at the zero separation etc.
  \subsection{Extended harmonic confinement model for hadron spectroscopy}
  Choice of scalar plus vector potential for the quark confinement has been successful in the predictions of the low lying hadronic properties in the relativistic  schemes for  the quark confinement \cite{kg1983,kg1987,jbp1996}. In relativistic harmonic confinement model (RHM), coloured quarks in a hadron  are confined through the action of a Lorentz scalar plus a vector harmonic  oscillator potential. It also leads to an alternate scheme for the
  understanding of nucleon-nucleon interactions from a more fundamental level \cite{vk1993}. The RHM has been extended to accommodate multiquark states from lighter to heavier flavour sectors with unequal quark masses \cite{JNPandya2001,PCV1999}.

  The mass of a hadron having $p$ number of quarks in this extended RHM (ERHM) is expressed as \cite{JNPandya2001,PCV1999},
  \beq
  M_N (q_1q_2.....)\ & = &\dis\sum_{i=1}^{p}\epsilon_N(q_i,p)_{conf}\ +\
  \dis\sum_{i<j=1}^{p}\epsilon_N (q_iq_j)_{coul} \nonumber\\ && + \dis\sum_{i<j=1}^p \epsilon_N^J(q_i,q_j)_{SD}    \label{eq:hadronmass}\eeq
  where the first sum is the total confined energies of the constituting quarks of the hadron,
  the second sum corresponds to the residual colour coulomb interaction energy between the confined quarks
  and the third sum is due to spin dependent terms.

  The intrinsic energies of the quarks using harmonic potential is given by
  \be
  \epsilon_N (q,p)_c = {\sqrt {(2N+3)\Omega_N (q) + M_q^2 - {3M_q\over{\dis\sum_{i=1}^{p}M_q(i)}} \Omega_0 (q)}} \label{eq:epsilon} \ee
  The coulombic part of the energy is computed using the residual coulomb potential given by \cite{gw1986},
  \be V_{coul}(q_iq_j) = \dis\frac{k \ \alpha_s(\mu)}{\omega_n r} \ee
  here $\omega_n$ represents the colour dielectric ``coefficient'' \cite{gw1986} which is found to be state dependent \cite{PCV1999}, so as to get consistent coulombic  contribution to the excited states of the hadrons. Such state dependence in the effective $\alpha_s$ for the excited states of quarkonia has been reported by others \cite{nb2001}. It is a measure of the confinement strength through the non-perturbative contributions to the confinement scale at the respective threshold  energy of the quark-antiquark excitations.

  The wave functions for quarkonia are constructed here by  retaining the nature of single  particle wave function but with a two particle size parameter $\Omega_N({q_iq_j})$  instead of $\Omega_N(q)$  \cite{JNPandya2001} and the coulomb energy is computed  perturbatively using the confinement basis with the two particle size parameter. The spin average (center of weight) masses of the $c\bar c$ and $b\bar b$ ground states are obtained by choosing the model parameters $m_c$~=~1.428~GeV,  $m_b$~=~4.637~GeV, $k = 0.19252$ and  the confinement parameter $A = 0.0685$ GeV$^{3/2}$ \cite{PCV1999}.

  From the centre of weight masses, the pseudoscalar and vector mesonic masses are
  computed by incorporating the residual two body chromomagnetic interaction through  the spin-dependent term of the COGEP perturbatively as,
  \be \epsilon_N^J(q_iq_j)_{S.D.}\ =\ \langle  NJ | V_{SD}|NJ \rangle  \ee
  where $|NJ\rangle $ is the given hadronic state. For mesons $|NJ\rangle $ becomes
  the $|q_i\bar q_j\rangle $ states.  We consider the two body spin-hyperfine
  interaction and the spin-orbit interaction of the residual (effective) confined one gluon exchange potential (COGEP) given by \cite{vvk1992},
  \beq
  V_{\sigma_i\cdot \sigma_j} &= & {\dis\frac{\alpha_s(\mu) N_i^2N_j^2}{4}}
  \dis\frac{\lambda_i\cdot \lambda_j}{[E_i + m_i] [E_j + m_j]} \nonumber\\
  && \times [4\pi \delta^3(r_{ij}) - C_{_{CCM}}^4 r^2
  D_1(r_{_{ij}})](-\frac{2}{3}{\sigma_i\cdot \sigma_j})
  \label{eq:ssintpot}\eeq
and  \beq V_{q_iq_j}^{LS} & = & \dis\frac{\alpha_s}{4}\cdot\dis\frac{N_i^2\
  N_j^2}{(E_i+M_i)(E_j+M_j)}\dis\frac{\lambda_i\cdot\lambda_j}{2r_{ij}} \nonumber\\
  && \times\left [4 \vec L \cdot \vec S \left (D_0'(r_{ij}) +
  2D_1'(r_{ij})\right ) \right ] \label{eq:sointpot1}\eeq
  Where $D_0'(r_{ij})$  and $D_1'(r_{ij})$ appeared in Eqns (\ref{eq:ssintpot}) and (\ref{eq:sointpot1})  are derivatives of the confined gluon propagators of CCM \cite{vvk1992} and $N_{i/j}$ are  the RHM normalisation constants given by,
  \be N_{i/j} = {\sqrt{2(E_{i/j}+M_{i/j})/(3E_{i/j} + M_{i/j})}} \ee
  Where, $C_{_{CCM}}$ corresponds to the confinement strength of the gluons and
  $r_{_{ij}}$ is the interquark distance. The computed masses for charmonia and bottomonia  based on this ERHM model are given in comparison with experimental and other theoretical model results in Tables  \ref{tab:ccmasses} and \ref{tab:bbmasses} respectively. The resultant spectroscopic masses and their radial wave functions of the quarkonia states will be employed for the study of their decay rates.
\begin{table}
\begin{center}
\caption {Masses of charmonia in GeV/c$^2$}
{\begin{tabular}{cccccccccccc} \hline\hline
  State   & ERHM  & \m{4}{c}{CPP$_\nu$} & \cite{pdg2006} & \cite{bgs2005}  & \cite{vij2007}  & \cite{efg2003} & \cite{bd2003} & \cite{rr2005}\\
  \cline{3-6}
  & &$_{\nu=0.5}$ & $_{\nu=1.0}$& $_{\nu=1.5}$& $_{\nu=2.0}$& PDG & &  &  & &\\
\hline
  $1^1S_0$ & 2.985 &3.000&2.950&2.912&2.882&2.980 & 2.981 & 2.990 & 2.979 & 3.093 & 2.982\\
  $2^1S_0$ & 3.626 &3.352&3.522&3.636&3.852&3.638 & 3.625 & 3.627 & 3.588 & 3.096 & 3.619\\
  $3^1S_0$ & 4.047 &3.541&3.912&4.212&4.436&--    & 4.032 & --   & 3.991 & --   & 4.053\\
  $4^1S_0$ & 4.424 &-    &   - & -   & -   &--    & 4.364 &  --  & --   & --   & -- \\
\hline
  $1^3S_1$ & 3.096 &3.092&3.112&3.129&3.144&3.097&3.089 & 3.097 & 3.096 & 3.096 & 3.097\\
  $2^3S_1$ & 3.690 &3.375&3.583&3.739&3.852&3.686&3.666 & 3.685 & 3.686 & 3.476 & 3.686\\
  $3^3S_1$ & 4.082 &3.553&3.950&4.285&4.547&4.040&4.060 & 4.050 & 4.088 & 3.851 & 4.102\\
  $4^3S_1$ & 4.408 & -   & -   &-    &-    &4.415 & 4.386 & 3.443 & --   & 4.223 & 4.447\\
\hline
  $1^3P_0$ & 3.431 &3.329&3.398&3.461&3.479&3.415 & 3.425 & 3.496 & 3.424 & 3.468 & 3.415\\
  $1^3P_1$ & 3.464 &3.302&3.424&3.504&3.556&3.511& 3.505 & 3.525 & 3.510 & 3.468 & 3.511\\
  $1^3P_2$ & 3.530 &3.323&3.477&3.590&3.673&3.556& 3.556 & 3.507 & 3.556 & 3.467 & 3.556\\
  $1^1P_1$ & 3.497 &3.313&3.450&3.547&3.615&3.526& 3.524 & --   & 3.526 & 3.467  & 3.524\\
\hline
  $2^3P_0$ & 3.891 &3.494&3.792&4.009&4.153&3.800& 3.851 & -- & 3.854  & 3.814 & 3.864\\
  $2^3P_1$ & 3.899 &3.507&3.832&4.085&4.269&3.880& 3.923 & -- & 3.929  & 3.815 & 3.950\\
  $2^3P_2$ & 3.916 &3.531&3.911&4.239&4.501&3.940& 3.970 & -- & 3.972  & 3.815 & 3.992\\
  $2^1P_1$ & 3.907 &3.519&3.872&4.162&4.385& --  & 3.941 & -- & 3.945  & 3.815 & 3.963\\
\hline\hline
  \end{tabular}}\label{tab:ccmasses}
  \end{center}
  \end{table}

\begin{table}
\begin{center}
  \caption{Masses of bottomonia in GeV/c$^2$}
  {\begin{tabular}{cccccccccccc}
  \hline\hline
  State   & ERHM  & \m{4}{c}{CPP$_\nu$} &\cite{pdg2006} & \cite{rac2006}  & \cite{rr2007}
  & \cite{efg2003} & \cite{jvij2005} & \cite{avs2003} \\
  \cline{3-6}
  & & $_{\nu=0.5}$ & $_{\nu=1.0}$ & $_{\nu=1.5}$ & $_{\nu=2.0}$ & PDG & & & & &\\
 \hline
  $1^1S_0$  & 9.425   &9.426&9.411&9.399&9.389&9.300&9.457&9.421&9.400  & 9.454 & 9.300\\
  $2^1S_0$  & 10.012  &9.696&9.826&9.924&9.995&--&10.018 & 10.004 & 9.993  & --   & 9.974\\
  $3^1S_0$  & 10.319  &9.824&10.088&10.334&10.529&--&10.380&10.350&10.328&--&10.333\\
  $4^1S_0$  & 10.572  &-  &-  &-  &-&--& 10.721 & 10.632 & --    & --   & --  \\
  $5^1S_0$  & 10.752  & - & - & - &-&--& 11.059 & --    & --    & --   & --  \\
 \hline
  $1^3S_1$  & 9.461   &9.463&9.468&9.472&9.475&9.460&9.460&9.460&9.460&9.505&9.460 \\
  $2^3S_1$  & 10.027&9.702&9.841&9.951&10.032&10.023&10.023&10.024&10.023&10.013&10.023\\
  $3^3S_1$  & 10.329&9.827&10.097&10.334&10.529&10.355&10.385&10.366&10.355&10.335&10.381\\
  $4^3S_1$  & 10.574  &-  &-  &-  && 10.579  & 10.727 & 10.643 &  --   & 10.577 & 10.787\\
  $5^3S_1$  & 10.753  & - & - & - &-& 10.865  & 11.065 & --    &  --   & 10.770 & 11.278\\
 \hline
  $1^3P_0$  & 9.839&9.664&9.755&9.820&9.866&9.859&9.894 & 9.860 & 9.863 & 9.855 & 9.865\\
  $1^3P_1$  & 9.873&9.670&9.775&9.820&9.866&9.893&9.941 & 9.892 & 9.892 & 9.875 & 9.895\\
  $1^3P_2$  & 9.941&9.683&9.792&9.866&9.913&9.912& 9.983 & 9.910 & 9.913 & 9.887 & 9.919\\
  $1^1P_1$  & 9.907&9.672&9.775&9.852&9.911&  --& 9.955  & 9.900 & 9.901 & --   & 9.894 \\
 \hline
  $2^3P_0$& 10.197&9.803&10.035&10.228&10.379&10.232&10.234&10.231&10.234&10.212&10.238\\
  $2^3P_1$  & 10.207&9.806&10.044&10.246&10.406&10.255&10.283&10.258&10.255&10.227& 10.264\\
  $2^3P_2$  & 10.227&9.811&10.062&10.282&10.462& 10.268  & 10.326  & 10.271 & 10.268 & 10.237 & 10.283\\
  $2^1P_1$  & 10.217&9.808&10.053&10.264&10.434&  --    & 10.296   & 10.263 & 10.261& --    & 10.260\\
  \hline\hline
\end{tabular}}
\label{tab:bbmasses}
\end{center}
\end{table}
\begin{table}
\begin{center}
\caption{Theoretical predictions of the ground state masses (in GeV) and $|R(0)|^2$
of heavy quarkonia}
{\begin{tabular}{lllccccc}
\hline\hline & Models &  $\alpha_s $& \m{2}{c}{$M_P$}&
\m{2}{c}{$M_V$} &$|R_{cw}(0)|^2$ \\
\cline{4-7}
&&&Theory&Expt \cite{particle2002}&Theory&Expt \cite{particle2002}&GeV$^3$\\
\hline
&ERHM \cite{JNPandya2001}&0.356&2.985&2.9804&3.096&3.097&0.556\\
&BT \cite{BuchmullerTye1981}$$&0.360&2.980&$\pm$&3.097&&0.810\\
$c \bar c$&PL(Martin) \cite{Martin1980}&0.430&2.980&0.0012&3.097&&0.999\\
&Log \cite{Quiggrosner1977}&0.370&2.980&&3.097&&0.815\\
&Cornell \cite{Eichten1978}&0.310&2.980&&3.097&&1.454\\
&$CPP_{\nu}$ \cite{AKRai2002}$\nu=$0.5&0.300&3.000&&3.092&&0.610\\
&\hspace{0.63in}=1.0&0.300&2.950&&3.112&&1.100\\
&\hspace{0.63in}=1.5&0.300&2.912&&3.129&&1.508\\
&\hspace{0.63in}=2.0&0.300&2.882&&3.144&&1.850\\
\hline
&ERHM \cite{JNPandya2001}&0.241&9.452&&9.464&&4.990\\
&BT \cite{BuchmullerTye1981}&0.241&9.377&9.300&9.464&9.4603&6.477\\
$b \bar b$&PL(Martin) \cite{Martin1980}&0.270&9.398&$\pm$&9.462&$\pm$&4.591\\
&LOG \cite{Quiggrosner1977}&0.245&9.395&0.002&9.460&0.00026&4.916\\
&Cornell \cite{Eichten1978}&0.217&9.335&$\pm$&9.476&&14.05\\
&$CPP_{\nu}$ \cite{AKRai2002}$\nu = $0.5&0.233&9.426&0.002&9.463&&3.908\\
&\hspace{0.63in}=1.0&0.233&9.411&&9.468&&5.988\\
&\hspace{0.63in}=1.5&0.233&9.399&&9.472&&7.728\\
&\hspace{0.63in}=2.0&0.233&9.389&&9.475&&9.181\\
\hline\hline
\end{tabular}}\label{tab:32}
\end{center}
\end{table}

\subsection{Coulomb plus power potential in variational approach}
The heavy-heavy bound state systems such as $c \bar c$, $b \bar b$, may also be described by a nonrelativistic Hamiltonian given by
 \cite{AKRai2002,AKRai2005,AKRai2006}
\begin{equation}
\label{eq:nlham}H=M+\frac{p^2}{2M_1}+V(r), \ee
where  $M = m_Q + m_{\bar Q}$, $M_1=\frac{m_Q \ m_{\bar Q}}{m_Q +
m_{\bar Q}}$, $m_Q$ and $m_{\bar Q}$ are the mass parameters, $p$ is
the relative momentum of each quark and $V(r)$ is the quark
antiquark potential given by \cite{AKRai2002} \be
V(r)=\frac{-\alpha_c}{r} + A r^\nu; \ \ \alpha_c=\frac{4}{3} \
\alpha_s \label{eq:405}\ee
Here, we use the Coulomb plus power potential where the power index
$\nu$ varies from 0.5 to 2. Each value of $\nu$ corresponds to
different potential scheme. Under the variation approach one requires
a trial wave function to compute the expectation value of the Hamiltonian. In our previous work \cite{AKRai2002,AKRai2005}, we have used the harmonic oscillator wave function for the study of the ground state masses, decay constants and decay rates. The value of the wave function at origin is very crucial parameter for estimation of decay constant and decay rates. But it is found that the radial wave function at r=0 are underestimated with the gaussian trial wave function. We also came across problem of overestimation with the Gaussian like wave function (HO basis) in the predictions of the orbital excited states. This suggests us to look for an alternate trial wave function for the present study.

Accordingly, we employ hydrogenic trial wave function given by
\be
\label{eq:wavfun}
R_{nl}(r) = \left(\frac{\mu^3 (n-l-1)!}{2n(n+l)!}\right)^{1/2} \
(\mu \ r)^l \ e^{- \mu r /2} \ L^{2l+1}_{n-l-1}(\mu
r)\ee
 for the variational calculations in the present study. Here, $\mu$ is the variational parameter and $L^{2l+1}_{n-l-1}(\mu r)$ is Laguerre polynomial. The Potential parameter, A for each choices of $\nu$ is fixed to yield the ground state mass of the meson.

The expectation value of the Hamiltonian for the ground
state is obtained as
 \be \label{eq:toteg}E(\mu,\nu )=M+\frac{1}{8}
 \frac{\mu^2}{M_1}+\frac{1}{2}\left(- \mu
\alpha_c + A \ \frac{\Gamma(\nu+3)} {\mu^{\nu}}\right)
\ee
The wave function parameter $\mu$ is determined here by using virial therom
for a chosen value of $\nu$. Thus, we obtain the spin average mass
$(M_{SA})$ of the system from Eqn (\ref{eq:toteg}). In our earlier study of quarkonium we have employed the gaussian trail wave function for the mass spectra, decay constant and decay rates, but the predictions of the decay constant and decay rates were underestimated significantly \cite{AKRai2002,AKRai2005}. For the same region in the present study we are using hydogenic like wave function. Unlike the earlier case of ERHM, here we consider the one gluon exchange for the spin-spin and spin-orbit interactions given by \cite{Gerstein1995}
\be \label{spindependent}
 V_{S_Q\cdot S_{\bar Q}}(r)= \frac{8}{9} \frac{ \alpha_s}{ m_Q m_{\bar Q}}
 {\vec S}_Q\cdot{\vec S}_{\bar Q} \ 4 \pi \delta(r)\ee and
\be \label{spinorbit}
  V_{L \cdot S}(r)= \frac{4 \ \alpha_s}{3 \ m_Q m_{\bar Q}}
  \frac{{\vec L}\cdot{\vec S}}{r^3}\ee
The value of the radial wave function $R(0)$ for $0^{- \ +}$ and
$1^{--}$ states would be different due to their spin dependent hyperfine interaction. The spin hyperfine interactions of the heavy flavour mesons are small and this can cause very little shift in the value of the wave function at the origin. Many models do not consider this contribution to the value of $R(0)$. However, we account this correction to the value of $R(0)$ by considering
\be R_{nJ}(0)=R(0)\left[1+(SF)_J \frac{\langle \varepsilon_{SD}\rangle _{nJ}}{M_{SA}} \right]\ee
Where $(SF)_J$ and $\langle \varepsilon_{SD}\rangle _{nJ}$ is the
spin factor and spin interaction energy of the meson in the $nJ$ state, while $R(0)$ and $M_{SA}$ correspond to the radial wave function at the zero separation and spin average mass respectively of the $Q \bar Q$ system. The mass parameters $m_b$ = 4.66 GeV, $m_c$ =1.31 GeV and the mass difference between the pseudoscalar and vector meson due to the chromomagnetic hyperfine interaction is computed as
described in Ref  \cite{AKRai2002}. The potential parameter $A$ of $CPP_{\nu A}$ is fixed at 0.276,  0.167, 0.101, 0.060  GeV$^{\nu+1}$ for $c\bar c$ and 0.195, 0.158, 0.126, 0.098 GeV$^{\nu+1}$ for $b\bar b$, for the choices of $\nu = 0.5,\ 1.0, 1.5, \ 2.0$ respectively to get the ground state spin average masses of the quarkonia. The mass spectra computed in this model are listed in tables \ref{tab:ccmasses} and \ref{tab:bbmasses} along with other model predictions. The computed ground state spectroscopic parameters like the vector and pseudoscalar masses, the square of the radial wave functions at the origin for $c \bar c$ and $b \bar b$ systems are tabulated in Table \ref{tab:32} for $\nu$ = 0.5, 1.0, 1.5 and 2.0. Similar parameters of other potential models are also listed. These data will be used to compute the decay rates.
\section{Decay rates of $c \bar c$ and $b \bar b$ mesons in NRQCD formalism}
\begin{table}
\begin{center}
\caption{$0^{-+} \rightarrow \gamma \ \gamma$ and $1^{-\ -}
\rightarrow l^+ \ l^-$ decay rates (in keV) of $c \bar c$ and
$b\bar b$ mesons.} {\begin{tabular}{clcccccc} \hline\hline
&Models&\m{3}{c}{$0^{-+} \rightarrow \gamma \ \gamma $}
&\m{3}{c}{$1^{-\ -} \rightarrow l^+ \ l^-$}\\
&&$\Gamma_o$&$\Gamma_{NRQCD}$&$\Gamma_{Expt.}$ \cite{particle2002}
& $\Gamma_{VW}$&$\Gamma_{NRQCD}$&$\Gamma_{Expt.}$ \cite{particle2002}\\
\hline &ERHM \cite{JNPandya2001}&7.67&3.401&&5.469&2.377&\\
&BT \cite{BuchmullerTye1981}&11.19&5.165&7.0$^{+ \ 1.0}_{- \ 0.9}$&8.31&3.711&5.40 $\pm$ 0.15\\
$c\bar c$&PL(Martin) \cite{Martin1980}&13.81&6.054&&9.96&4.691&$\pm $0.07\\
&Log \cite{Quiggrosner1977}&11.26&5.364&&8.13 &3.942&\\
&Cornell \cite{Eichten1978}&20.09&8.614&&14.50&6.704&\\
&$CPP_{\nu}$ \cite{AKRai2002}$\nu =$0.5&8.173&2.213&&6.130&1.766&\\
&\hspace{0.78in}=1.0&14.649&5.513&&11.053&3.017&\\
&\hspace{0.78in}=1.5&19.971&7.889&&15.165&3.938&\\
&\hspace{0.78in}=2.0&24.297&9.920&&18.549&4.591&\\
\hline
&ERHM \cite{JNPandya2001}&0.440&0.207&&1.320&0.620&\\
&BT \cite{BuchmullerTye1981}&0.569&0.278&&1.717&0.833&\\
$b\bar b$&PL(Martin) \cite{Martin1980} &0.408&0.195&0.364 \cite{Bodwin1995}&1.216&0.588&1.314 $\pm$ 0.029\\
&Log \cite{Quiggrosner1977}&0.437&0.211&&1.303&0.633&\\
&Cornell \cite{Eichten1978}&1.258&0.597&&3.719&1.801&\\
&$CPP_{\nu}$ \cite{AKRai2002}$\nu = $0.5&0.345&0.163&&1.035&0.489&\\
&\hspace{0.78in}=1.0&0.529&0.251&&1.587&0.748&\\
&\hspace{0.78in}=1.5&0.683&0.324&&2.047&0.965&\\
&\hspace{0.78in}=2.0&0.811&0.385&&2.433&1.146&\\
\hline\hline
\end{tabular}}\label{tab:34}
\end{center}
\end{table}
\begin{table}
\begin{center}
 \caption{Leptonic decay widths (in keV) of $c\bar c(n^3S_1)$}
{ \begin{tabular}{ccccccccccc}
  \hline\hline
  State & ERHM &\m{2}{c}{CPP$_\nu$}&\cite{pdg2006} & \cite{rr2007} & \cite{rr2007}
  & \cite{qr1979} & \cite{eit1978} & \cite{smi2006} \\
  \cline{3-4}
  & & $\nu=0.5$&$\nu=1.0$ &PDG &Pert &Non-pert &&&\\
 \hline
  $J/\psi(1^3S_1)$ & 5.469 &6.130&11.053&5.55 $\pm$ 0.14 & 4.28 & 1.89 & 4.80 & 7.82 & 6.72 $\pm$ 0.49 \\
  $\psi(2^3S_1)$& 2.140 &1.232&2.973&2.48 $\pm$ 0.06 & 2.25 & 1.04 & 1.73 & 3.83 & 2.66 $\pm$ 0.19 \\
  $\psi(3^3S_1)$& 0.796 &0.551&1.495&0.86 $\pm$ 0.07 & 1.66 & 0.77 & 0.98 & 2.79 & 1.45 $\pm$ 0.07 \\
  $\psi(4^3S_1)$& 0.288 &-&-& 0.58 $\pm$ 0.07 & 1.33 & 0.65&0.51&2.19 & 0.52 $\pm$ 0.02 \\
  \hline\hline
  \end{tabular}}\label{tab:ccleptdcaywdth}
  \end{center}
  \end{table}
\begin{table}
\begin{center}
\caption{Leptonic decay widths (in keV) of $b\bar b(n^3S_1)$}
\begin{tabular}{cccccccccc} \hline\hline
  State & ERHM &\m{2}{c}{CPP$_\nu$}& \cite{pdg2006} & \cite{smi2006} & \cite{vva2007} &\cite{rac2006} & \cite{rr2007} & \cite{rr2007} \\
\cline{3-4}
&  &$\nu=0.5$&$\nu=1.0$& PDG & &  &  & Pert & Non-pert \\
\hline
  $\Upsilon(1^3S_1)$ & 1.320&1.035&1.587&1.340 $\pm$ 0.018 & 1.45 $\pm$ 0.07 & 1.314 & -- & 5.30 & 1.73 \\
  $\Upsilon(2^3S_1)$ & 0.628&0.155&0.390& 0.612 $\pm$ 0.011 & 0.52 $\pm$ 0.02 & 0.576 & 0.426 & 2.95 & 1.04 \\
  $\Upsilon(3^3S_1)$ & 0.263&0.066&0.211& 0.443 $\pm$ 0.008 & 0.35 $\pm$ 0.02 & 0.476 & 0.356 & 2.17 & 0.81 \\
  $\Upsilon(4^3S_1)$ & 0.104 &-&-& 0.272 $\pm$ 0.029 & -- & 0.248 & 0.335 & 1.67 & 0.72 \\
  $\Upsilon(5^3S_1)$ & 0.040 &-&-& 0.310 $\pm$ 0.070 & -- & 0.310 & 0.311 & -- & -- \\
\hline\hline
\end{tabular}\label{tab:bbleptdcaywdth}
\end{center}
\end{table}
The decay rates of the heavy quarkonium states, light hadrons, photons or pairs of leptons are among the earliest applications of perturbative QCD \cite{app,de,Barbieri1979}. The decay rates of the mesons are
factorized into a short-distance part that is related to the annihilation rate of the heavy quark and antiquark and a long-distance part containing all nonperturbative effects of the QCD. The short-distance factor calculated in terms of the running coupling constant $\alpha(M)$ of QCD is evaluated at the scale
of the heavy-quark mass $M$, while the long-distance factor is expressed in terms of the meson's nonrelativistic wave function, or its derivatives, evaluated at origin.  The di-gamma decay of $^1S_0$ state and the leptonic decay of $1^{--}$ state using the conventional Van-Royen Weisskopf formula \cite{Vanroyenaweissskopf}
\be \label{eq:0gamma}
 \Gamma_0 = \frac{12 \alpha_e^2 e_Q^4}{M_P^2} \ R^2_P(0) \ee and
\be
\label{eq:vwgamma} \Gamma _{VW}= \frac {4 \alpha_e^2 e_Q^2}{M_V^2}
\ R^2_V(0) \ee
as well as using the NRQCD formalism have been computed. The NRQCD factorization expressions
for the decay rates are given by \cite{Bodwin1995}
 \beq \label{eq:nq1}
\Gamma(^1 S_0 \rightarrow \gamma \gamma)&=&\frac{2 Im f_{\gamma \gamma}(^1S_0)}{m^2_Q} \left |\langle 0|\chi^{\dag}\psi|^1 S_0\rangle \right|^2 \nonumber\\ && + \frac{2 Im g_{\gamma \gamma}(^1 S_0)}{m^4_Q} \nonumber \\ && \times Re \left [\langle ^1 S_0|\psi^{\dag}\chi|0\rangle \langle 0|\chi^{\dag}(-\frac{i}{2}\overrightarrow{D})^2\psi|^1 S_0\rangle \right] \nonumber\\ && + O(v^4 \Gamma ) \eeq
\beq \label{eq:nq2}
\Gamma(^3 S_1 \rightarrow e^+e^-)&=&\frac{2 Im f_{ee}(^3
S_1)}{m^2_Q} \left |\langle 0|\chi^{\dag}\sigma \psi|^3 S_1\rangle \right|^2 \nonumber\\ && + \frac{2 Im g_{ee}(^3 S_1)}{m^4_Q} \nonumber\\ && \times Re \left [\langle ^3 S_1|\psi^{\dag}\sigma\chi|0\rangle \langle 0|\chi^{\dag}\sigma(-\frac{i}{2}\overrightarrow{D})^2\psi|^3 S_1\rangle \right] \nonumber\\ && + O(v^4 \Gamma ) \eeq
The matrix elements that contributes to the decay rates of the S wave states into $\eta_c\rightarrow \gamma \gamma$ and $\psi \rightarrow e^+e^-$ through next-to-leading order in $v^2$, the vacuum-saturation approximation gives \cite{Bodwin1995}
\be
\langle ^1S_0|{\cal{O}}(^1 S_0)|^1S_0\rangle  =  \left |\langle 0|\chi^{\dag}\psi|^1S_0\rangle \right|^2[1 + O(v^4 \Gamma)]\ee
\beq
\langle ^3S_1|{\cal{O}}(^3 S_1)|^3S_1\rangle =\left |\langle 0|\chi^{\dag}\sigma \psi|^3S_1\rangle \right|^2[1 + O(v^4 \Gamma)]\eeq
\beq
\langle ^1S_0|{\cal{P}}_1(^1 S_0)|^1S_0\rangle & = & Re \left [\langle ^1S_0|\psi^{\dag}\chi|0\rangle \langle 0|\chi^{\dag}(-\frac{i}{2}\overrightarrow{D})^2\psi|^1S_0\rangle \right] \nonumber\\ && + O(v^4 \Gamma ) \eeq
\beq
\langle ^3S_1|{\cal{P}}_1(^3 S_1)|^3S_1\rangle & = &
Re \left [\langle ^3S_1|\psi^{\dag}\sigma\chi|0\rangle \right. \nonumber\\ && \left. \langle 0|\chi^{\dag}\sigma(-\frac{i}{2}\overrightarrow{D})^2\psi|^3S_1\rangle \right] \nonumber\\ && + O(v^4 \Gamma ) \eeq
The Vacuum saturation allows the matrix elements of some four fermion operators to be expressed in terms of the regularized wave-function parameters given by   \cite{Bodwin1995}
\be
\langle ^1S_0|{\cal{O}}(^1 S_0)|^1S_0\rangle =\frac{3}{2
\pi}|R_{^1S_0}(0)|^2[1+O(v^4)]\ee
\be
\langle ^3S_1|{\cal{O}}(^3 S_1)|^3S_1\rangle =\frac{3}{2
\pi}|R_{^3S_1}(0)|^2[1+O(v^4)]\ee
\be
\langle ^1S_0|{\cal{P}}_1(^1 S_0)|^1S_0\rangle
 = -\frac{3}{2 \pi}|\overline{R^*_{cw}}\ \overline{\bigtriangledown^2 R_{cw}}|[1+O(v^4)] \ee
\be
\langle ^3S_1|{\cal{P}}_1(^3 S_1)|^3S_1\rangle =-\frac{3}{2 \pi}|\overline{R^*_{cw}}\ \overline{\bigtriangledown^2 R_{cw}}|[1+O(v^4)] \ee
The factorization formula for electromagnetic annihilation, the decay rates for $\eta_c\rightarrow \gamma \gamma$ and $\psi \rightarrow e^+e^-$ are
\beq \label{eq:sogma}
\Gamma(^1 S_0 \rightarrow \gamma \gamma)&=&\frac{2Im f_{\gamma \gamma}(^1
S_0)}{\pi\ m^2_Q} \left |R_{^1S_0}\right|^2 \nonumber\\ && - \frac{N_c\ Im g_{\gamma \gamma}(^1 S_0)}{\pi m^4_Q} Re (\overline{R_s^*} \ \overline{\nabla^2 R_s}) \nonumber\\ && + O(v^4 \Gamma ) \eeq
 \beq
\Gamma(^3 S_1 \rightarrow e^+e^-)&=&\frac{N_c\ Im f_{ee}(^3
S_1)}{\pi\ m^2_Q} \left |R_{^3 S_1}\right|^2 \nonumber\\ && - \frac{N_c\ Im g_{ee}(^3 S_1)}{\pi m^4_Q} Re (\overline{R_s^*} \ \overline{\nabla^2 R_s}) \nonumber\\ && + O(v^4 \Gamma ) \eeq
The short distance coefficients f's and g's computed in the order
of $\alpha^2$ as \cite{Bodwin1995}
\be Im f_{\gamma \gamma}(^1 S_0)= \pi Q^4 \alpha^2\ee
\be Im g_{\gamma \gamma}(^1S_0)=- \frac{4 \pi Q^4}{3} \alpha^2 \ee
\be Im f_{ee}(^3 S_1) = \frac{\pi Q^2 \alpha^2}{3} \ee
\be Im g_{ee}(^3 S_1)=- \frac{4 \pi Q^2}{9} \alpha^2 \ee
Where $C_F=(N_c^2-1)/(2N_c)=4/3$.

The decay rates are computed using conventional Van-Royen Weisskopf formula for $\Gamma_0$ $\&$ $\Gamma_{VW}$ as well as using NRQCD expressions. Here $ N_c=3$ is the numbers of colour
and $\alpha$ is the electromagnetic coupling constant. The
phenomenological mass parameters and the strong running coupling
constant employed in the meson mass predictions are being employed
here to compute the decay rates. For  comparison, we have listed the
results in Tables \ref{tab:34}, \ref{tab:ccleptdcaywdth} and
\ref{tab:bbleptdcaywdth}.
\section{Summary and Conclusion}
We have presented the decay rates of the charmonia and bottomonia systems using spectroscopic parameters obtained from different potential models within the NRQCD formalism. As the non-perturbative aspect of the interaction has already been taken into account in the confinement schemes of quarks and gluons, the residual one gluon exchange effects employed in this study are treated perturbatively. The predictions of harmonic plus coulomb potential (ERHM)are found to reproduce the experimental results with good accuracy. The predicted $\eta_c(2S)$ mass is in good agreement with recent experimental mass.

Using the radial wave functions of $c \bar c$ and $b \bar b$ systems of different potential models,  the decay rates of $0^{-+}\ra\gamma\gamma$ and $1^{--}\ra l^+l^- $ are computed within the NRQCD formalism. The results are compared with the values obtained using the conventional formula ($\Gamma_0$, $\Gamma_{VW}$) as well as with the respective experimental results. It is observed from Table \ref{tab:34} that the computed radiative decay of $\eta_c$ from ERHM using the conventional formula ($\Gamma_0$) is closer to the experimental value while all other model predictions using NRQCD formalism found to be closer to the experimental values  within its error bars on either side. Similar is the case for the predictions of the $\Gamma_{ee}$ width of $J/\psi$. The leptonic decay widths of higher excited states also found to be in good agreement with experimental values. Further, the radiative decay rates of $\eta_b$ meson predicted by all the models except Cornell potential are close to each other. In case of $b\bar b$, all the models seem to be in good agreement with experimental results.\\

Though the predictions using conventional formula are far from the experimental results in most cases, the predictions based on NRQCD are found to be in accordance with the experimental results. The present study in the determination of the $S$ wave masses and decay rates of $c\bar c$ and $b\bar b$ systems provide future
scopes to study light hadron decays and various transition between the excited states of these mesonic systems \cite{Brambilla2005}. It can be concluded that the NRQCD formalism has most of the corrective contributions required for most of the potential models for the right predictions of the decay rates. It must also be noted that the ERHM model predictions of the decay rates are in excellent agreement with the experimental values. It may be due to the fact that its parametric description of the quarkonia through the spectroscopy is more close to reality.\\

In case of CPP$_\nu$, no single value for $\nu$ is found to account for all decay widths and masses correctly. The purpose of varying the potential index $\nu$ is to observe the variation of the different properties of the quarkonia with the choice of where different potential forms. From the present study it is also found that the potential index $\nu$ is not the only factor affecting estimation of masses and decay rates as the wave function at $r=0$ is an important parameter. The values of the decay rates are overestimated with hydrogenic wave function while largely underestimated with gaussian trial wave function \cite{AKRai2005}. Hence, the present study indicates that the right form of the wave function lies between the hydrogenic to gaussian i.e. of the form $e^{-\mu r^p}$ with $1<p<2$.

The present study also suggests that the spectroscopic parameters perhaps are not sufficient for understanding the decay dynamics of hadrons.\\

{\bf Acknowledgement:}\\
PCV acknowledges the financial support from Department of Science and Technology, Government of India under a Major research project SR/S2/HEP-20/2006.


\begin{thebibliography}{99}
\bibitem{nb2007}            N. Brambilla, NRQCD and Quarkonia, arXiv:hep-ph/0702105v2 (2007).
    \bibitem{M.B.Voloshin2008}M. B. Voloshin; hep-ph/0711.4556v3 (2008)
\bibitem{Eichten2008}E. Eichten, S. Godfrey, Hanna Mahlke and Jonathan L. Rosner;  hep-ph/0701208v3 (2008)
\bibitem{BuchmullerTye1981} W. Buchm\"uller and S. H. H. Tye, {\it Phys.Rev.} {\bf D 24}, 132 (1981).
\bibitem{Martin1980}        A. Martin, {\it Phys. Lett.} {\bf B 93}, 338 (1980).
\bibitem{Quiggrosner1977}   C. Quigg and J. L. Rosner, {\it Phys. Lett.} {\bf B 71}, 153 (1977).
\bibitem{Eichten1978}       E. Eichten {\it et. al.}, {\it Phys. Rev.}  {\bf D 17}, 3090 (1978).
\bibitem{AKRai2002}         A. K. Rai, R. H. Parmar and P. C. Vinodkumar, {\it Jnl. Phys.} {\bf G 28}, 2275 (2002).
\bibitem{AKRai2005}         A. K. Rai, J. N. Pandya and P. C. Vinodkumar, {\it Jnl. Phys.} {\bf G 31}, 1453 (2005).
\bibitem{AKRai2006}         A. K. Rai and P. C. Vinodkumar, {\it Pramana J. Phys.} {\bf 66}, 953 (2006), hep-ph/0606194.
\bibitem{Gerstein1995}      S. S. Gershtein, V. V. Kiselev, A. K. Likhoded and A. V. Tkabladze, {\it Phys. Rev.} {\bf D 51}, 3613 (1995).
\bibitem{Hafsakhan1996}     H. Khan and P. Hoodbhoy,  {\it Phys.  Rev.} {\bf D 53}, 2543 (1996).
\bibitem{Bodwin1995}        G. T. Bodwin, E. Braaten and G. P. Lepage, {\it Phys. Rev.} {\bf D 51}, 1125 (1995); {\bf D 55}, (E) ( 1997).
\bibitem{Bodwin2006}        G. T. Bodwin, D. Kang and J. Lee, {\it Phys. Rev.} {\bf D 74}, 014014 (2006);
\bibitem{Bodwin2008}        G. T. Bodwin, H.S. Chang, D. Kang, J. Lee and Chaehyun Yu, {\it Phys. Rev.} {\bf D 77}, 094017 (2008);
\bibitem{JNPandya2001}      J. N. Pandya and P. C. Vinodkumar, {\it Pramana J. Phys.} {\bf 57}, 821 (2001).
\bibitem{PCV1999}           P. C. Vinodkumar, J. N. Pandya, V. M. Bannur and S. B. Khadkikar,{\it Eur. Phys. J.} {\bf A 4}, 83 (1999).
\bibitem{kg1983}            S. B. Khadkikar and S. K. Gupta, Phys. Lett. {\bf B 124}, 523  (1983).
\bibitem{kg1987}            S. K. Gupta and S. B. Khadkikar, Phys. Rev. {\bf D 36}, 307 (1987).
\bibitem{jbp1996}           Jena, Behera and Panda, Phys. Rev. {\bf D 54}, 11 (1996).
\bibitem{vk1993}            K. B. Vijaya Kumar and S. B. Khadkikar, Nucl. Phys. {\bf A 556},
        396  (1993).
\bibitem{gw1986}            K. Gottfried and V. F. Weisskopf, {\it Concepts of Particle Physics}, 397 (1986).
\bibitem{nb2001}  N. Brambilla, Y. Sumino and A. Vairo, {\it Phys. Lett.} {\bf B 513}, 381-390 (2001).
\bibitem{vvk1992}           P. C. Vinodkumar, K. B. Vijaya Kumar and S. B. Khadkikar, {\it Pramana Jnl. of Phys.} {\bf 39}, 47 (1992).
\bibitem{app}               T. Appelquist and H. D. Politzer, {\it Phys. Rev. Lett.} {\bf 34}, 43 (1975).
\bibitem{de}                A. De Rujula and S. L. Glashow, {\it Phys. Rev.} {\bf D 38}, 46 (1975).
\bibitem{Barbieri1979}      R. Barbieri, G. Curci, E. d'Emilio and E. Remiddi; {\it Nucl. Phys. } {\bf B 154}, 535 (1979).
\bibitem{Vanroyenaweissskopf} R. Van Royen and V. F. Weisskopf, {\it Nuovo Cimento} {\bf 50}, 617 (1967).
\bibitem{Bodwin2002}        G. T. Bodwin and A. Petrelli, {\it Phys. Rev.} {\bf D 66}, 094011 (2002).
\bibitem{Brambilla2005}     N. Brambilla {\it et. al.}, {\it CERN report:}  487 (2005); hep-ph/0412158.
\bibitem{particle2002}      S. Eidelman {\it et. al.}, {\it Phys. Lett.} {\bf B 592}, 1 (2004).
\bibitem{pdg2006}           Y. M. Yao {\em et al} (Particle Data Group); {\it J. Phys.} {\bf G 33}, 1 (2006).
\bibitem{bgs2005}           T. Barnes, S. Godfrey and E. S. Swanon, {\it Phys. Rev.} {\bf D 72}, 054026 (2005).
\bibitem{vij2007}           J. Vijande, N. Barnea and A. Valcarce, {\it Int. J. Mod. Phys.} {\bf A 22}, 561 (2007).
\bibitem{efg2003}           D. Ebert, N. Faustov and V. O. Galkin,  {\it Phys. Rev.} {\bf D 67}, 014027 (2003).
\bibitem{bd2003}            A. E. Bernardini and C. Dobrigkeit, {\it J. Phys.} {\bf G 29}, 1439 (2003).
\bibitem{rr2005}            S. Radford and W. W. Repko, {\it Int. J. Mod. Phys.} {\bf A 20}, 3774 (2005).
\bibitem{jvij2005}          J. Vijande, F. Fernandez and A. Valcarce; {\it J. Phys.} {\bf G 31}, 481 (2005).
\bibitem{smi2006}           S. M. Ikhdair, R. Sever; {\it Int. J. Mod. Phys.} {\bf A 21}, 3989 (2006).
\bibitem{vva2007}           V.V. Anisovich, L.G. Dakhno, M.A. Matveev, V.A. Nikonov and A.V. Sarantsev; {\it Phys. Atom. Nucl.} {\bf 70}, 63 (2007).
\bibitem{rac2006}           R. A. Coimbra and O. Oliveira; hep-ph/0610142 (2006).
\bibitem{rr2007}            S. F. Radford, W. W. Repko; Phys. Rev. {\bf D 75},  074031 (2007).
\bibitem{avs2003}           A. V. Shoulgin, G. M. Vereshkov, O. V. Lutchenko; J. Phys. {\bf G 29}, 1245 (2003).
\bibitem{aeb2003}           A. E. Bernardini and C. Dobrigkeit; J. Phys. {\bf G 29}, 1439 (2003).
\bibitem{eit1978}           E. Eichten, K. Gottfried, T. Kinoshita, K. D. Lane and T. M. Yan; Phys. Rev. {\bf D 17}, 3090 (1978).
\bibitem{qr1979}            C. Quigg and J. L. Rosner, Phys. Reports {\bf 56},  222 (1979).
\end{thebibliography}
\end{document}